\documentclass[12pt]{aastex}

\shorttitle{Migration and MHD Turbulence} 
\shortauthors{Steinacker, Laughlin, \& Adams}

\newcommand{\be}{\begin{equation}}
\newcommand{\ee}{\end{equation}}
 
\newcommand{\twig}{ {\tilde t} } 
\newcommand{\fifty}{ {\cal F}_{50} } 
\newcommand{\frms}{ {\langle {\cal F}_{50} \rangle} } 
\newcommand{\sigdisk}{ \sigma_{\rm d} } 
\newcommand{\ts}{{ t_{\rm smp} }} 
\newcommand{\porb}{{ P_{\rm orb} }} 
\def\kms{\ifmmode \hbox{ \rm km s}^{-1} \else{ km s$^{-1} $}\fi}

\begin{document}

\title{Type I Planetary Migration with MHD Turbulence} 

\author{Gregory Laughlin,$^1$, Adriane Steinacker,$^2$ and 
Fred C. Adams$^3$}  
 
\affil{$^1$Lick Observatory, University of California, Santa Cruz, CA 95064} 

\affil{$^2$NASA Ames Research Center, Moffett Field, CA 94035} 

\affil{$^3$Michigan Center for Theoretical Physics, 
University of Michigan, Ann Arbor, MI 48109}

\begin{abstract} 

This paper examines how type I planet migration is affected by the 
presence of turbulent density fluctuations in the circumstellar disk.
For type I migration, the planet does not clear a gap in the disk and its
secular motion is driven by torques generated by the wakes it creates
in the surrounding disk fluid.  MHD turbulence creates additional
density perturbations that gravitationally interact with the planet
and can dominate the torques produced by the migration mechanism itself.
This paper shows that conventional type I migration can be readily 
overwhelmed by
turbulent perturbations and hence the usual description of type I
migration should be modified in locations where the magnetorotational
instability is active. In general, the migrating planet does not
follow a smooth inward trend, but rather exhibits a random walk
through phase space. Our main conclusion is that MHD turbulence will alter
the time scales for type I planet migration and -- because of chaos --
requires the time scales to be described by a distribution of values. 

\end{abstract}


\section{Introduction} 

Ever since their initial discovery (Mayor \& Queloz 1995, Marcy \&
Butler 1996), extrasolar planets have been found in unexpectedly
short-period orbits.  Since theories of planet formation suggest that
planets must form further out in the disk (e.g., Lissauer 1993), these
findings strongly indicate the necessity of planetary migration (e.g.,
Lin, Bodenheimer, \& Richardson 1996). A variety of migration scenarios 
have been suggested. Type I migration (Ward 1997) occurs when a planet 
(or protoplanet) is embedded in a circumstellar gaseous disk and the
planet's mass is too small to clear a gap in the disk. In this case,
the planet drives wakes into the background fluid of the disk and
these wakes, in turn, exert torques on the planet. These torques are
generally greater in the inward direction than in the outward
direction, and a net inward migration occurs.

These same disks are expected to contain magnetic fields that are
subject to a robust instability (Balbus \& Hawley 1991), for regimes
in which an ideal-MHD description is adequate. The nonlinear outcome
of this magneto-rotational instability is MHD-driven turbulence. Indeed,
MHD-driven turbulence is commonly invoked as a primary source 
of disk viscosity which can drive mass accretion through the disk.
The usual scenario for type I migration, however,
implicitly assumes a smooth background, uninterrupted by turbulence.
Given that such turbulence, along with associated stochastic density
fluctuations in the disk gas may often be present, we
re-examine the type I migration scenario by considering the dynamical
evolution of embedded protoplanets that are torqued by transient density
perturbations arising from MHD turbulence. 

The effects of large-scale
turbulence on type II migration, where the planet clears a gap in the
disk (e.g., Goldreich \& Tremaine 1980; Lin \& Papaloizou 1993), has
been explored recently (Nelson \& Papaloizou 2003; Papaloizou \&
Nelson 2003; Winters, Balbus, \& Hawley 2003).  In a related vein,
Terquem (2003) has considered the possibility of stopping inward
migration by a strong toroidal magnetic field. These previous papers
show how turbulence and magnetic fields can produce important
modifications to the standard migration picture. This present work
carries out a complementary analysis for type I migration.

Our approach to the problem is as follows. We first compute a representative
three-dimensional MHD simulation (\S 2) in a circumstellar disk and
use the results to quantify the spectrum of nonlinear density perturbations
that are produced by MHD-driven turbulence. These density
perturbations provide a stochastic time-varying gravitational torque on 
any embedded protoplanets that are orbiting in the disk. Our next step is to
construct a parametric model which characterizes the essential
aspects of the spectrum
of turbulent density fluctuations that are observed in the three-dimensional
simulation. This model is described in terms of a time-varying
non-axisymmetric potential function(\S 3).  The resulting prescription for the
perturbations is then applied to two-dimensional hydrodynamical
simulations of circumstellar disks to study type I protoplanetary 
migration (\S 4).
With this scheme, we can examine the effects of the turbulent
perturbations on the migration of bodies that are not sufficiently
massive to open gaps in a protostellar disk, and we can obtain
an overall initial survey of the Type I migration problem. Our approach 
allows for a rapid survey of the important regions of parameter space,
and is complementary to full-scale 3D MHD simulations which trace the
evolution of a planet in an ionized circumstellar disk.
Because the
saturation of the magnetorotational instability occurs at nonlinear
amplitudes, MHD-driven turbulence naturally produces
large density fluctuations, which in turn produce torques that
generally dominate those produced by the planetary wakes of
objects with $M<10M_{\oplus}$. We quantify this threshold by deriving 
an analytic
estimate of the fluctuation amplitudes that are required for type I migration
to be highly distorted (\S 5) and then discuss the longer term 
behavior (\S 6). 

\section{MHD Simulations} 

Our first step is to conduct a representative three-dimensional
simulation to evaluate the expected properties of MHD turbulence in a
model cicumstellar disk.  We use the ``Nirvana'' MHD code
(Ziegler \& R\"udiger 2000) as modified by Steinacker \& Papaloizou
(2002). We work in cylindrical coordinates ($z$,$R$,$\phi$), and use
a system of units in which the gravitational constant, $G$, and the
central stellar mass, $M$, are both unity.
Our disk model has a nearly Keplerian radial domain that extends 
from an inner boundary at
$R$=1.5 out to $R$=3.5, and covers an azimuthal slice that subtends
$\pi/3$ radians. The equation of state is taken to be locally isothermal, with
$P(R)=\rho(R)c_{s}(R)^{2}$
(see e.g. Laughlin \& Bodenheimer 1994). This approximation is reasonable
for optically thick disks whose local radiative diffusion time is
shorter than the local dynamical (orbital) timescale.
The gravitational potential from the
central star is specified to depend only on the cylindrical radial
coordinate $\Phi=-GM/R$. 
The model therefore does not consider the vertical density gradient
that results from the $z$ component of the gravitational force. This
approximation simplifies the physical situation, allowing the use
of a modest number of zones in the vertical direction. An obvious
future refinement will be to study vertically stratified models. The reference
simulation described here uses 334 radial zones, 72 azimuthal zones, and 40
vertical zones. 

In the nearly Keplerian region of our initial
equilibrium model, the disk density varies inversely with
radius, 
$\rho(R)=\rho_{0}(R_{0}/R)$,
the soundspeed is given by 
$c_{s}(R)=C_{0}\sqrt{R_{0}/R}$,
and the rotational velocity is 
$v_{\phi}(R)=\sqrt{GM/R_{0}-2C_{0}^{2}}\sqrt{R_{0}/R}$. For numerical
reasons, the disk also 
contains inner and outer radial boundary domains as described in 
Steinacker \& Papaloizou (2002). 

We adopt $C_{0}=0.1$, along with a subthermal poloidal seed field having
zero net flux through the disk. With these starting parameters,
the disk experiences
the magnetorotational instability, progresses through the channel
phase, and develops turbulence as described in detail by Steinacker \&
Papaloizou (2002). This configuration of initial conditions was found by 
Steinacker and Papaloizou (2002) and Hawley (2001) (among others) to
lead to the smallest overall amplitude in the turbulent density fluctuations.
Initial magnetic configurations with net flux threading the disk can 
lead to turbulent fluctuations that are up to two orders of magnitude larger.
This turbulence persists for the duration of the
simulation, which was followed to 200 orbits at the outer disk edge.

Figure 1 shows a map of the surface density distribution $\sigma(R,\phi)$ in
the disk after the turbulence has entered its quasi-steady fully
developed phase. The order-unity surface density fluctuations that are
present in the
disk provide the stochastic gravitational torques the are capable of
severely modifying the type I migration process. The corresponding
global Fourier amplitudes
\be
C_{m}={ \vert \int_{0}^{2\pi} \int_{R_{in}}^{R_{out}}
\sigma(r,\phi)r\,dr \exp^{-im\phi}d\phi \vert \over{
\int_{0}^{2\pi}\int_{R_{in}}^{R_{out}}\sigma(r,\phi)rdrd\phi}} \, ,
\label{global}
\ee
of the disk fluctuations shown in Figure 1 are plotted
as the filled black circles in Figure 2. The perturbation amplitude
is strongest
for the lowest azimuthal wavenumber ($m=6$) that fits in the
$\pi/3$ azimuthal domain. As the azimuthal mode number $m=6n$ is
increased through n=2,3,4, and 5, the Fourier amplitudes show a gradual
decline. The concentration of power in lower-order Fourier modes
is a characteristic of MHD-generated turbulence, and has been previously 
observed by other authors, including Armitage (1998), and Hawley et al (1995).

\section{Parameterization of the Turbulence} 

In order to survey the effects of MHD turbulence on type I planet
migration, we have chosen to construct a simple heuristic model to
describe the perturbations. By characterizing the turbulence arising
in the full MHD treatment in a simple manner, we can use the resulting
prescription to specify the development of time-varying disk
fluctuations
in two-dimensional planet migration simulations. The alternative (and
more rigorous) approach to the problem would involve running an expensive
three-dimensional MHD simulation with every planet migration
calculation. Each long-duration run of the MHD code requires,
for example, 30,000 CPU hours on an Origin 3800 parallel processing
facility (Nelson \& Papaloizou 2003). Running enough
simulations to sample the parameter space of interest is thus overly 
time-consuming.

We are primarily interested in the gravitational forces exerted on a
migrating planet by the density perturbations arising from the
MHD instabilities.  We thus need a description of the potential
perturbations $\Phi$, which can be differenced to compute forces that
act on the fluid elements; the gravitational forces from the resulting
density enhancements then act on the planet itself.  Specifically, we
adopt a simple heuristic model for the potential $\Phi$ induced by the 
turbulent perturbations: 
\be
\Phi = {A \xi {\rm e}^{-(r-r_c)^2/\sigma^2} \over r^{1/2}} \,
\cos\bigl[ m \theta - \varphi - \Omega_c \twig) \bigr] \, 
\sin\bigl[ \pi {\twig \over \Delta t} \bigr] \, . 
\label{eq:modeform} 
\ee
With enough modes of this form, one can characterize the spectrum of
density fluctuations produced by the MHD turbulence.  Given the finite
resolution of our codes, the computed MHD perturbations can be
reasonably modeled by applying 50 modes at any given time. The modes come in and
out of existence with the time dependence specified above, where
$\twig$ = $t - t_0$.  An individual mode begins at time $t_0$ and has faded away
after a time $\Delta t$ = $t_f - t_0$. After the mode is gone, a new
mode is generated. With the proper choice of the distributions of the
remaining parameters (see below), we can reasonably mimic the power spectrum
of density fluctuations in
the actual MHD turbulence computed by the full three-dimensional
code (see Figure 2).

The following parameter choices are found to work: The modes are
centered at radial location $r_c$ and initial angular location (phase
angle) $\varphi$.  The location of the mode $r_c$ is chosen from a
random (uniform) distribution, as is the phase angle $\varphi$. The
extent of the mode is determined by the azimuthal wave number $m$
which is chosen to be distributed according to a log-random
distribution for wavembers in the range $2 \le m \le N{grid}/8$
(where $N_{grid}$ is the number of azimuthal zones in the simulation,
typically 512 here). We find that the choice of a log-random distribution of $m$
gives the best approximation to the density fluctuation spectrum produced
by the MHD simulation; this spectrum of $m$ provides power
on all spatial scales. With
$m$ specified, the mode extends for a distance $2 \pi r_c /m$ along the
azimuthal direction. The radial extent is then specified by choosing
$\sigma = \pi r_c/4m$ so that the mode shapes have roughly a 4:1 aspect
ratio (this statement is not precise because the profiles differ in
the radial and azimuthal directions).  Notice also that the initial
mode shapes are not tilted as would be expected from the usual spiral
patterns seen in disks; here, the durations of the modes are generally
much longer than the time required for the Keplerian shear to adjust
the orientations, so we can rely on the shear to take over.

The modes first appear at a time $t_0$, which is chosen so that the
simulation contains 50 active modes at all times.  The pattern speed
$\Omega_c = v_{kepler}/r_c$ in the time dependent factor allows the mode
center to travel along with the Keplerian flow. The temporal duration
of the mode $\Delta t$ = $t_f - t_0$ is taken to be the sound
crossing time of the mode along the angular direction, i.e., 
$\Delta t$ = $2 \pi r_c/(m a_S)$. 

Finally, the modes have an overall amplitude $A$ and an individual
amplitude $\xi$. As written, the amplitude $A$ has units of $[A] =
\ell^{5/2} t^{-2}$ (where $\ell$ is a length scale and $t$ is a time
scale) so that the overall expression has units of potential. The
dimensionless variable $\xi$ has a gaussian distribution with unit
width, whereas the overall amplitude $A$ has a fixed value for all of
the modes.  We can find the amplitude $A$ that reproduces the level of
perturbations found in the MHD simulation. Alternately, we can treat
the amplitude $A$ as an adjustable parameter and study the effects of
these turbulent fluctuations on planet migration as a function of the
amplitude (see below).

Figure 2 shows the power spectra from both the full
MHD treatment of the previous section and the parametric model
developed in this section. The two spectra are in good agreement. As a
result, we employ this approach to specify the perturbations
for migration calculations, as presented in the following section.

\section{Migration with Turbulent Perturbations} 

With the perturbations due to MHD turbulence specified (as described
in \S 3), we have performed fluid dynamic simulations of the migration
of a terrestrial planet (or core of a giant planet) in the
circumstellar disk. These simulations use a two-dimensional fluid
dynamics code where we add the turbulent perturbations according to
the prescription described above.

Our two-dimensional code
(Laughlin 1994) is based on second-order
van Leer advection (e.g. Stone \& Nornam 1992). We generally use a grid of
256 evenly spaced radial zones and 512 azimuthal zones. The initial model
has the same surface density and sounspeed distributions and the same active
radial domain as was employed in the three-dimensional simulation. We 
minimize reflection by implementing sponge boundary conditions at
the radial edges of the computational domain. In the
${n_{d}}_{i}=12$ radial zones interior to $\xi=1.6$, and in the
${n_{d}}_{o}=12$ radial zones exterior to $\xi=3.4$, we smoothly impose
an admixture of the initial equilibrium solution 
into the hydrodynamically computed variables $x_{\rm hydro}(\xi)$.
That is, at the inner edge we have,
\be
x(\xi_{j})=({j-1 \over{ {n_{d}}_{i}-1 }}) x_{\rm hydro}(\xi_{j}) +
({ {{n_{d}}_{i}-j} \over{ {n_{d}}_{i}-1 }}) x_{\rm equil}(\xi_{j}) \, .
\ee
This damping is applied at a cadence ${t_{d}}_{i}=\Delta\xi({n_{d}}_{i})/c_{g}$
at the inner edge, and ${t_{d}}_{o}=\Delta\xi({n_{d}}_{o})/c_{g}$ at the outer
edge, where $\Delta\xi(j)$ is the radial width of the uniformly spaced zones.

The planet mass is taken to be either 1 or 10 $M_{\oplus}$, and is
initially placed in a circular
orbit of semi-major axis (radius) $a$ = 2.5 AU. (Note that all of
these radii can be scaled). Over this restricted annulus 
covered by the hydrodynamical domain, the initial disk
mass set to be a small fraction of the stellar mass, i.e., $M_{\rm d} /
M_\ast$ = 0.02.  The subsequent evolution of a 
$1 M_{\oplus}$  planet in the $a-e$
plane is shown in Figure 3 for three different amplitudes of the turbulent
perturbations: $A=5.0\times10^{-6}$, $A=5.0\times10^{-5}$,
and $A=5.0\times10^{-4}$. The corresponding surface density perturbations are
depicted in Figures 4, 5, and 6.

For small fluctuation amplitudes $A$, type I migration proceeds in
largely unperturbed fashion. The planet slowly and steadily spirals
inward. The fluid dynamic simulations only follow the planet for
$\sim100$ orbits in this class of simulations (see Figs. 3a and 4a).
When, on the other hand, the fluctuation
amplitude is large, the chaotic perturbations produced indirectly
through turbulence dominate the torques produced by steady type I
migration. In this regime, the evolution of the planet is highly
chaotic (see Figs. 3c and 4c).  In between these extremes, there
exists a critical fluctuation amplitude, the value of $A$ at which
type I migration changes its character, but a general inward migration
trend is still evident (Figs. 3b and 4b). The critical amplitude
determined empirically from these two-dimensional disk simulations is
$A_E \sim 1 \times 10^{-5}$.  In the following section, we derive an
analytic estimate for this critical value.

\section{The Critical Amplitude} 

Equation (\ref{eq:modeform}) gives the potential produced by the
turbulent fluctuations. The gradient of this potential provides a
forcing term that acts on the gas in the disk (but does not act
directly on the planet) and thereby produces corresponding
fluctuations in the surface density of the disk. To obtain a rough
understanding of the critical fluctuation amplitude $A_C$ required for
turbulence to overwhelm the standard type I migration torques, we
assume here that the gravitational potential of the perturbed gaseous
disk (where this potential does act on the planet) has the same form
as that of equation (\ref{eq:modeform}). However, the amplitude of the
potential acting on the planet will, in general, be reduced from that
acting on the gas by a factor $\Gamma \le 1$ (which we estimate
below).  The torques exerted on the planet by the surface density
perturbations in the disk will thus have the form 
\be
\tau = - {m A \xi \Gamma M_P \over r^{1/2} } \, 
{\rm e}^{-(r-r_c)^2/\sigma^2} 
\sin [ m \theta - \varphi - \Omega_c \twig ] 
\sin [ \pi \twig / \Delta t ] \, , 
\ee
where $M_P$ is the mass of the planet and all the other quantities are
defined previously.  At any given time, the disk will contain 50
modes that exert torques of this form. To evaluate the effectiveness
of these torques, we need to define an average torque strength. First, 
we integrate over time to obtain the net change in angular momentum 
per mode, i.e., 
\be
\Delta J = \int_0^{\Delta t} \, dt \, \tau \, = 
{m A \xi \Gamma M_P \over r^{1/2} } 
\, {\rm e}^{-(r-r_c)^2/\sigma^2} \, {\Delta t \over \pi} 
\, \, {\sin(a\pi - \varphi) - \sin\varphi \over a^2 - 1} \, , 
\ee
where the parameter $a$ is defined to be $a \equiv (m \Omega -
\Omega_c) \Delta t / \pi$.  The mean torque per mode $\tau_1$,
averaged over the lifetime of the mode, is thus given by
\be
\tau_1 = {\Delta J \over \Delta t} = 
{A \Gamma M_P \over r^{1/2} } {\cal F} \, 
\qquad {\rm where} \qquad  
{\cal F} \equiv {m \xi \over \pi} \, \, {\rm e}^{-(r-r_c)^2/\sigma^2} 
\, \, {\sin(a\pi - \varphi) - \sin\varphi \over a^2 - 1} \, , 
\ee
where the variables $r_c$, $\varphi$, $\xi$, and $m$ are chosen 
randomly as described in \S 3. 

At any given time, the disk contains 50 modes and hence 50 values of
the function $\cal F$. Since these torques can either be positive or
negative, the 50 modes will tend to cancel each other out, and the net
effect forcing effect will be relatively small. We thus need to
compute the forcing function $\fifty$ defined by 
\be
\fifty \equiv \sum_{j=1}^{50} {\cal F}_j \, , 
\ee
where the ${\cal F}_j$ are sampled in the same way that the numerical
simulations sample parameter space. The function $\fifty$ should
average to zero (or some low value) because the torques can be
positive or negative.  However, the RMS value of $\fifty$, denoted
here as $\frms$, should provide a good measure for the effective
strength of this torque. Numerically sampling the function shows that
$\frms \approx 0.29$. Putting these results together, we find that 
the average torque $\tau_T$ exerted on the planet through the action 
of turbulence is given by 
\be 
\tau_T = \frms {A \Gamma M_P \over r^{1/2} } \approx 
0.29 {A \Gamma M_P \over r^{1/2} } \, . 
\label{eq:torqueT} 
\ee 

To estimate the response amplitude $\Gamma$ of the disk (due to the
potential induced by turbulent fluctuations), we use a simple
linearized WKB treatment of a two-dimensional disk (Shu 1992). Unlike
the usual spiral density wave theory, however, we include only the
forcing due to the fluctuation potential $\Phi$ and neglect the
self-gravity of the disk. In the WKB limit, we can solve for the radial
part of the reduced surface density perturbation $S$ in terms of the
forcing potential $\Phi$ to obtain 
\be 
S = {k^2 \sigdisk \Phi \over (\omega - m \Omega)^2 - 
\kappa^2 - k^2 a_S^2} \, , 
\label{eq:sresult} 
\ee 
where the epicylic frequency $\kappa$ = $\Omega$ for a Keplerian disk.
Next, we need to use the relationship between the surface density
perturbation $S$ and the corresponding gravitational potential $V$.
In the WKB limit, this relation has been calculated previously (e.g.,
Shu 1992) and can be written in the form 
\be 
S \approx - { |k| V \over 2 \pi G} \, . 
\label{eq:wkb} 
\ee 
Equating the expressions for the surface density response to the
turbulent fluctuations (eq. [\ref{eq:sresult}]) and the corresponding
perturbation in the gravitational potential (eq. [\ref{eq:wkb}]), we
can solve for the response factor $\Gamma$, i.e., 
\be
\Gamma \approx {V \over \Phi} \approx 
{2 \pi G \sigdisk k \over (\omega - m \Omega)^2 - 
\Omega^2 - k^2 a_S^2} \, . 
\ee
To evaluate this expression, we must make further simplifying
assumptions. The turbulent fluctuations induce modes with relatively
large wavenumbers $m$ = 2 -- 32, so we can approximate the
denominator as $| {\cal D} | \sim m^2 \Omega^2$ = $m^2 G M_\ast/r^3$.
Given the radial dependence of the turbulent fluctuations
(eq. [\ref{eq:modeform}]), we can estimate the radial wavenumber $k$
in the WKB limit: $kr = (r/\Phi) (\partial \Phi/\partial r) \sim 2 r^2
/ \sigma^2$ $\sim 32 m^2 / \pi^2$. The expression for $\Gamma$ thus
becomes 
\be
\Gamma = {\pi \sigdisk r^2 \over M_\ast } {64 \over \pi^2} 
\, \approx 0.06 \, \, . 
\ee 

Next, we need to compare the average torque produced through turbulent
fluctuations with that produced by the usual type I migration
mechanism (Ward 1997). This torque $\tau_I$ can be written 
\be 
\tau_I = \beta_I  (M_P/M_\ast)^2 (\pi \sigdisk r^2) 
(r \Omega)^2 (r/H)^3 \, , 
\label{eq:torqueI} 
\ee
where $H$ is the disk scale height, $\sigdisk$ is the disk surface
density, and $\beta_I$ is a dimensionless amplitude. Previous
calculations suggest that $\beta_I \sim 10^{-2}$ (Ward 1997).

By equating $\tau_T$ and $\tau_I$, we can solve for the critical
amplitude $A_C$ for which the torques due to turbulence become larger
than those of standard type I migration. We find 
\be
A_C = {17 \pi^2 \over 64} \beta_I (M_P/M_\ast) (r/H)^3 
r^{1/2} (r \Omega)^2 \, \approx 1 \times 10^{-5} \, ,  
\label{eq:acrit} 
\ee
where the numerical value is obtained using the same parameters as in
the numerical simulations presented here (for a one Earth mass planet).  
This analytic estimate for the critical amplitude is in good agreement
with that observed in the numerical simulations themselves -- which 
show an empirically determined value of $A_E \sim 1 \times 10^{-5}$ 
(see \S 4). 

Before leaving this section, we note that the torque produced by type
I migration (eq. [\ref{eq:torqueI}]) and that produced indirectly
through turbulence (eq. [\ref{eq:torqueT}]) scale differently with the
mass of the planet. Small planetesimals, with masses $m \ll M_E$,
will be tossed around violently because of MHD turbulence. On the
other hand, sufficiently large planets can be impervious to its
effects (for sufficiently small amplitudes $A$).  If we assume that
the maximum fluctuation amplitude $A_{\rm max} \approx$ $r^{1/2} 
(r \Omega)^2$, then we can define a critical mass scale $M_{CP}$, 
such that larger masses will always be dominated by type I migration 
torques. This mass scale is given by $M_{CP}/M_\ast$ = $(64/\pi^2
\beta_I)$ $(r/H)^3$ $\approx$ 0.04. Secondary bodies with such large
masses are generally brown dwarfs and not planets.  Furthermore, such
large bodies tend to open up gaps and experience type II migration
(Goldreich \& Tremaine 1980; Lin \& Papaloizou 1993; Ward 1997). As a
result, there is no regime of parameter space for which type I
migration torques necessarily dominate those produced by MHD
turbulence.

\section{Long Term Evolution} 

Given our analytic description of the effects of turbulence, we can
understand the possible range of migration behavior over longer time
intervals. This analytic approach is necessary, as it is difficult for
numerical simulations to remain stable over millions of orbits, the
dynamical time frames of the migration process. As we discuss below,
several complications arise in the long term.

First, even if the amplitude of the turbulent fluctuations is that
required for the resulting torques to have the same amplitude as those
of type I migration (i.e., if $A$ = $A_C$), the two sources of torque
have different accumulative effects. The type I migration torque
$\tau_I$ is a secular torque that consistently moves the planet
in one direction. The net turbulent torque $\tau_T$ can be either positive or
negative and changes its value over a sampling time that we denote
here at $\ts$. The changes in angular momentum accumulate in random
walk fashion.  The change in angular momentum over the integrated time
interval $t = N \ts$, for the two contributions, can thus be written
in the form 
\be 
(\Delta J)_I = N \tau_I \ts \qquad {\rm and} \qquad 
(\Delta J)_T = \sqrt{N} \tau_T \ts \, . 
\ee 
The sampling time $\ts$ is the time required for the disk to produce
an independent realization of the turbulent torques. The lifetime of
a mode with wavenumber $m$ is $\Delta t$ = $2 \pi r / (m a_S)$, which
can be written in the form $\Delta t = (2 \pi / \Omega) (r/mH)$. The
first factor is the orbital period $\porb$. The second factor is of
order unity: The disk generally has $r/H \approx 10$ and the
wavenumber $m$ lies in the range $2 \le m \le 32$. Since the
wavenumbers $m$ are logarithmically spaced and the lowest $m$ modes
live the longest (and are near the peak of the power spectrum -- see
Fig. 2), the appropriate value of $m$ to use in estimating the
sampling time lies near the low end of the range. As a result, we
expect $\ts \sim \porb$.  In order for the turbulent fluctuations to
dominate over a specified time interval $t_f$, the amplitude of the
turbulent fluctuations must be larger than the critical value $A_C$ by
an additional factor, i.e., 
\be 
A \ge \Bigl( {t_f \over \porb} \Bigr)^{1/2} A_C \, . 
\label{eq:acc} 
\ee 

A sufficiently large amplitude can keep the turbulent fluctuations
dominant over the entire lifetime of the disk. The expected disk
lifetime $t_{\rm disk}$ (specifically, the time over which planet
migration takes place) is of order 1 -- 3 Myr (Haisch, Lada, \& Lada
2001), while expected timescales for Type I migration of individual
(Earth mass) planetesimals are
considerably shorter, of order $10^{5}$ years
(Ward 1997).  The orbit time $\porb$ is of order 1 --3 yr. So this problem
contains another critical amplitude, namely that required for the
turbulent fluctuations to overwhelm type I migration torques over a
$10^{5}$ year secular migration timescale.
  This second critical amplitude $A_2$ is estimated to be 
\be
A_2 \approx \Bigl( {t_{\rm disk} \over \porb} \Bigr)^{1/2} A_C \, 
\sim 0.003 \, . 
\ee  

A useful benchmark is the number of steps $N$ (or, equivalently, 
the time scale $N \porb$) required for the random walk process to 
completely change the angular momentum. In other words, we set 
$(\Delta J)_T$ = $\sqrt{G M_\ast r}$ in equation (\ref{eq:acc}) 
and solve for $N$ to find 
\be 
N = \Bigl[ \Bigl( {A \over r^{5/2} \Omega^2} \Bigr) \, \, 
2 \pi \fifty \Gamma \Bigl]^{-2} \, . 
\ee 
Even for fluctuations with the critical amplitude $A=A_C$, this number
is rather large: $N \sim 3 \times 10^{8}$, with a corresponding time
scale of 300 Myr. On the other hand, for $A=A_2$, the required number
of steps falls to $N \sim 3 \times 10^3$, with a time scale of $\sim
3000$ yr.

Another complication is the radial dependence of the critical
amplitude $A_C$. As shown by equation (\ref{eq:acrit}), this ratio
scales as $A_C \propto r^{3 - 3q/2}$, where $q$ is the power-law index
of the disk temperature profile. Although the temperature will not be
purely a power-law, $q$ is nonetheless expected to lie the range $1/2
\le q \le 3/4$, so that the index $3-3q/2$ = 15/8 $-$ 9/4 $\approx$ 2. 
As a result, the amplitude of the turbulent modes required to dominate
the type I migration torques is a rapidly increasing function of
radius $r$ (decreasing as the planet moves in).  As the planet
migrates inward, $A_C \propto r^2$ decreases and turbulence tends to
dominate to an ever greater extent. On the other hand, if MHD
turbulence shuts down in the inner disk due to magnetic decoupling
(e.g., Fromang, Terquem, \& Balbus 2002), then standard type I
migration is resumed.

\section{Discussion and Conclusion} 

In this paper, we have generalized the scenario of type I planet
migration to include the effects of fluctuations produced by MHD
turbulence. To characterize the turbulence itself, we have run
three-dimensional MHD simulations to study the onset and character of
the turbulence, and used the results to specify the power spectrum of
turbulent fluctuations.  These perturbations were then parameterized
and used in a second set of two-dimensional numerical simulations that
study the migration phase itself. Finally, we have developed an
analytic description of the turbulent fluctuations, the torques they
produce, and their long term effects on the migration process. Our
specific results are summarized below:

MHD turbulence produces a full spectrum of fluctuations that peaks at
relatively low azimuthal wavenumbers $m \sim 1 - 6$ (Figs. 1 and 2).
These fluctuations can be characerized by potential fluctuations of
the form given by equation (\ref{eq:modeform}). The formulation
developed here may be useful in many other astronomical
applications because it allows the fluctuations produced by turbulence
to be effectively modeled with a minimum of computational effort, and 
even allows for many results to be determined analytically (see \S 5). 

The effects of turbulence on type I migration depend sensitively on
the amplitude of the turbulent potential perturbations.  Further, we
have found the critical threshold $A_C$ for this amplitude, using both
two-dimensional numerical simulations of the migration process
(Figs. 3 and 4) and an analytic treatment (eq. [\ref{eq:acrit}]).  
For perturbation amplitudes below this threshold ($A < A_C$) type I
migration proceeds smoothly inward, as envisioned by the original
scenario. For small but finite amplitudes, the turbulence leads to
random fluctuations superimposed on the smooth inward migration, but
the general trend remains. If the amplitude exceeds the threshold
($A>A_C$), the perturbations due to turbulence produce torques that
are stronger than those induced by the planetary wake, and the
evolution changes dramatically. Instead of displaying a smooth, inward
progression, the planet exhibits random walk behavior. The planet can
even move outward when the turbulence amplitude is large enough.

The critical amplitudes are $A_C \approx 1 \times 10^{-5}$ for a 1
Earth mass planet and $A_C \approx 1 \times 10^{-4}$ for a 10 Earth
mass planet. These critical amplitudes are found both in the
two-dimensional numerical simulations of type I migration and by the
analytic treatment of \S 5.  For comparison, the expected amplitudes
for the perturbations are $A \sim Z \sim 10^{-3}$, comfortably larger
than the critical amplitudes needed to change the migration scenario.
As a result, we expect MHD turbulence to change the character of the
type I migration process under typical disk conditions.

The main conclusion of this paper is that the typical migration
torques in a circumstellar disk will often be dominated by the
perturbations due to MHD turbulence rather than the steady planetary wakes
that drive type I migration. As a result, type I migration will often
display far richer behavior than has been considered previously. When
MHD instabilities are robust, migration will generally proceed in a
highly chaotic fashion, with an element of a random walk behavior,
although the underlying conventional type I secular torque acts to 
drive the planet inwards over
the long term. The complications introduced by turbulence have two
important implications: (1) The time scale for planet migration can be
quite different, either longer or shorter, when turbulence is present.
(2) The result of any given migration episode in a circumstellar disk
cannot be described by a single outcome; instead, due to chaos and
extreme sensitivity to initial conditions, the result of a migration
episode must be described in terms of a distribution of possible
outcomes. In particular, the time scale required for planets to
migrate inward will display a full distribution of values.

This chaotic element -- including the necessity of finding the full
distribution of possible outcomes -- has recently been emphasized for
migration scenarios that involve planet-scattering (Adams \& Laughlin
2003).  Other recent work (Nelson \& Papaloizou 2003; Papaloizou \&
Nelson 2003; Winters, Balbus, \& Hawley 2003) outlines the effects of
turbulence and chaos for type II migration.  Since this paper shows
that type I migration can often be dominated by turbulence/chaos as
well, nearly all possible mechanisms for planet migration are expected
to display chaotic behavior.  The ubiquity of chaotic processes
during the formation and early evolution of planetary systems certainly
contributes to the startling diversity that is observed among the known
extrasolar planets

\medskip 
 
This work was supported by NASA Ames through an award from the
Director's Discretionary Fund program, and by the NASA Origins
of Solar Systems Program under contract No. RTOP 344-37-22-12.

\newpage 
\begin{figure}
\figurenum{1} 
\epsscale{1.0}
%\plotone{fig1.ps}
\caption{Surface density distribution in a $\pi/3$ wedge of a circumstellar 
disk resulting from a three-dimensional simulation of MHD turbulence.
Gravitational torques arising from these fluctuations will stochastically
alter the orbits of protoplanets embedded in the disk.}
\end{figure}  

\newpage 
\begin{figure}
\figurenum{2} 
\epsscale{.80}
%\plotone{fig2.ps} 
\caption{Non-axisymmetric modal 
amplitudes observed in the numerical simulations. The 
solid black circles show the global fourier amplitudes $C_{6m}$ measured
in the MHD-turbulent three-dimensional simulation at the time plotted in
Figure 1. The solid red triangles show the amplitudes $C_{m}$
observed in a two-dimensional 
hydrodynamical simulation in which a $1M_{\oplus}$ protoplanet
is embedded in a 0.02 $M_{\odot}$ disk, and perturbed as described by 
equation 2 with $A=5.0\times10^{-4}$.
The open red squares show the amplitudes $C_{m}$ observed in a similar simulation with
$A=5.0\times10^{-5}$. The solid blue circles
show the amplitudes $C_{m}$ observed in a simulation containing
a $1M_{\oplus}$ protoplanet embedded in a disk with $A=0$.}
\end{figure} 

\newpage 
\begin{figure}
\figurenum{3} 
\epsscale{1.00} 
%\plotone{fig3.ps}  
\caption{Type I migration simulations using a two-dimensional fluid
code and the parametric treatment of the perturbations due to MHD
turbulence. The three panels show time evolution of the migrating
planet in the $a-e$ plane for fluctuations with low amplitude (a) where 
$A=5.0\times10^{-6} \sim A_C/2$, threshold amplitude (b) where 
$A=5.0\times10^{-5} \sim 5 A_C$, and high amplitude
(c) where $A=5.0\times10^{-4} \sim 50 A_C$. 
All of these simulations use a planet with one
Earth mass. Planets with higher mass behave similarly if the amplitude
of the turbulent fluctuations is increased accordingly (see text). }
\end{figure}  

\newpage 
\begin{figure}
\figurenum{4} 
\epsscale{1.00} 
%\plotone{fig4.ps}  
\caption{Type I migration simulation using a two-dimensional fluid
code and the parametric treatment of the perturbations due to MHD
turbulence. Perturbations in the surface
density of the disk are shown after 50 orbits of an embedded $1 M_{\oplus}$
protoplanet for the case of turbulent fluctuations with low
amplitude $A=5.0\times10^{-6} \sim A_C/10$. 
The color scale ranges from
$\sigma(r,\phi)/<\sigma(r)>$=0.998 (black) to $\sigma(r,\phi)/
<\sigma(r)>$=1.00836 (orange).}
\end{figure}

\newpage 
\begin{figure}
\figurenum{5} 
\epsscale{1.00} 
%\plotone{fig5.ps}  
\caption{
Same as the previous figure, but with 
amplitude $A=5.0\times10^{-5} \sim A_C$. The color scale ranges from
$\sigma(r,\phi)/<\sigma(r)>$=0.992 (black) to $\sigma(r,\phi)/
<\sigma(r)>$=1.00861 (orange).}
\end{figure}  

\newpage 
\begin{figure}
\figurenum{6} 
\epsscale{1.00} 
%\plotone{fig6.ps}  
\caption{
Same as the previous figure, but with 
amplitude $A=5.0\times10^{-4} \sim 10A_C$. The color scale ranges from
$\sigma(r,\phi)/<\sigma(r)>$=0.899 (black) to $\sigma(r,\phi)/
<\sigma(r)>$=1.0801 (orange).}
\end{figure}  

\end{document}